\documentclass[10pt,conference]{IEEEtran}
\usepackage{balance}
\usepackage{listings}
\usepackage{xcolor, colortbl}
\usepackage{graphicx}
\usepackage{booktabs}
\graphicspath{ {./res/images/} }
\usepackage{subcaption}
\usepackage{comment}
\usepackage{multirow}
\usepackage{wrapfig}
\usepackage{amsmath}
\usepackage{amssymb}
\usepackage{amsthm}
\usepackage{multicol}
\usepackage{hyperref}
\usepackage[linesnumbered,lined,commentsnumbered,ruled]{algorithm2e}
\def\BibTeX{{\rm B\kern-.05em{\sc i\kern-.025em b}\kern-.08em
    T\kern-.1667em\lower.7ex\hbox{E}\kern-.125emX}}

\theoremstyle{definition}
\newtheorem{definition}{Definition}[section]

\newcommand{\tool}{\textsc{flack}}
\newcommand{\bgd}[1]{\ifnum\value{lstnumber}=#1\color{lightgray}\fi}

\definecolor{keywords}{RGB}{0,35,197}
\definecolor{comments}{RGB}{19,121,0}
\definecolor{highlight}{RGB}{255,255,102}

\lstdefinelanguage{Alloy}
    {
    basicstyle=\small\sffamily, 
    morekeywords={and,let,Int,then,check,module,run,sig,abstract,extends,one,lone,some,set,open,module,assert,not,fact,pred,all,no,fun,in,implies,else,no},
	keywordstyle=\color{black}\bfseries,
	numbers=left,					
	numberstyle=\scriptsize,			
	stepnumber=1,					
	numbersep=7pt,
	backgroundcolor=\color{white},	
	showspaces=false,				
	showstringspaces=false,			
	showtabs=false,					
	frame=b, 
	tabsize=2,						
	captionpos=b,					
	breaklines=true,					
	breakatwhitespace=false,			
	numberbychapter=false,
	aboveskip=0.25em,
	belowskip=0.25em,
	escapeinside={(*@}{@*)},
	xleftmargin=1.5em,
	escapeinside={(*@}{@*)},
        keywordstyle=\color{keywords}\bfseries,
       commentstyle=\color{comments},
        morecomment=[l][\color{comments}]{//},
        morecomment=[l][\color{comments}]{--},
       stringstyle=\color{strings}
}

\newcommand{\code}[1]{\lstinline[language=Alloy]{#1}}

\begin{document}

\title{{\tool}: Counterexample-Guided Fault Localization for Alloy Models}

\author{
	\IEEEauthorblockN{Guolong Zheng\IEEEauthorrefmark{1}, ThanhVu Nguyen\IEEEauthorrefmark{1}, Simón Gutiérrez Brida\IEEEauthorrefmark{2}, Germán Regis\IEEEauthorrefmark{2}, \\Marcelo F. Frias\IEEEauthorrefmark{3}, Nazareno Aguirre\IEEEauthorrefmark{2}, Hamid Bagheri\IEEEauthorrefmark{1}}
    \IEEEauthorblockA{\IEEEauthorrefmark{1}Univeristy of Nebraska-Lincoln
    \\\{gzheng, tnguyen\}@cse.unl.edu, bagheri@unl.edu}
    \IEEEauthorblockA{\IEEEauthorrefmark{2}University of Rio Cuarto and CONICET
    \\\{sgutierrez, gregis, naguirre\}@dc.exa.unrc.edu.ar
		\IEEEauthorblockA{\IEEEauthorrefmark{3}Dept. of Software Engineering Instituto Tecnológico de Buenos Aires
		\\mfrias@itba.edu.ar}
		}
}

\maketitle

\begin{abstract}
Fault localization is a practical research topic that helps developers identify code locations that might cause bugs in a program.
Most existing fault localization techniques are designed for imperative programs (e.g., C and Java) and rely on analyzing correct and incorrect executions of the program to identify suspicious statements.
In this work, we introduce a fault localization approach for models written in a declarative language, where the models are not ``executed,'' but rather converted into a logical formula and solved using backend constraint solvers. We present {\tool}, 
a tool that takes as input an Alloy model consisting of some violated assertion and returns a ranked list of suspicious expressions contributing to the assertion violation. The key idea is to analyze the differences between \emph{counterexamples}, i.e., instances of the model that do not satisfy the assertion, and instances that do satisfy the assertion to find suspicious expressions in the input model.
The experimental results show that {\tool} is efficient (can handle complex, real-world Alloy models with thousand lines of code within 5 seconds), accurate (can consistently rank buggy expressions in the top 1.9\% of the suspicious list), and useful (can often narrow down the error to the exact location within the suspicious expressions).

\end{abstract}

\thispagestyle{plain}
\pagestyle{plain}

\section{Introduction}

Declarative specification languages and the corresponding formally precise analysis engines have long been utilized to solve various software engineering problems. The Alloy specification language~\cite{alloy} relies on first-order relational logic, and has been used in a wide range of applications, such as program verification~\cite{tacoconf}, 
test case generation~\cite{testcase,ICSE2016}, 
software design~\cite{DBLP:conf/icse/BagheriS04,DBLP:conf/gpce/BagheriS12,DBLP:conf/kbse/BagheriSS10},
network  security~\cite{network1,network2,network3}, security analysis of emerging platforms, such as IoT~\cite{iot_issta} and Android~\cite{android,android2}, and design tradeoff analysis~\cite{trademaker_ICSE,trademaker_TSE_2017}, to name a few. Cunha and Macedo, among others, use a recent extension of Alloy, called Electrum~\cite{DBLP:conf/kbse/BrunelCCM18}, to validate the European Rail Traffic Management System, a system of standards for management and inter-operation of signaling for the European railways~\cite{alloyuse3}. Kim~\cite{alloyuse7} proposes a  Secure Swarm Toolkit (SST), a platform for building an authorization service infrastructure for IoT systems, and uses Alloy to show that SST provides necessary security guarantees. 

Similar to developing programs in an imperative language, such as C or Java, developers can make subtle mistakes when using Alloy in modeling system specifications, especially those that capture complex systems with non-trivial behaviors, rendering debugging thereof even more arduous. 
These challenges call for debugging assistant mechanisms, such as fault localization techniques, that support declarative specification languages.

However, there is a dearth of fault localization techniques developed for Alloy. 
AlloyFL~\cite{alloyfl} is perhaps the only fault localization tool available for Alloy as of today.
The key idea of AlloyFL is to use ``unit tests," where a test is a predicate that describes an Alloy instance to encode expected behaviors, to compute suspicious expressions in an Alloy model that fails these tests.
To compute the suspicious expressions, AlloyFL uses mutation testing~\cite{mut1,mut2} and statistical debugging techniques~\cite{spfl1,spfl2,spfl3}, i.e., it mutates expressions, collects statistics on how each mutation affects the tests, then uses this information to assign suspicion scores to expressions.




  

While AlloyFL pioneered fault localization in the Alloy context and the obtained results thereof are promising, it relies on the assumption of the availability of AUnit tests~\cite{aunit}---i.e., predicates representing Alloy instances---which are not common in the Alloy setting. Indeed, instead of writing test cases, Alloy users write assertions to describe the desired property and let the Alloy Analyzer search for potential counterexamples (cex's) that violate the property.
Moreover, it is unclear how many test cases are needed or how good they must be for AlloyFL to be effective (e.g., in the AlloyFL evaluation~\cite{alloyfl}, the number of tests ranges from 30 to 120).

To address this state of affairs and improve the quality of Alloy development, 
we present an automated approach and an accompanying tool-suite for \textbf{f}ault \textbf{l}ocalization in \textbf{A}lloy models using \textbf{c}ounterexamples, called {\tool}. 
Given an Alloy model and a property that is not satisfied by the model, {\tool} first queries the underlying Alloy Analyzer for a counterexample, an instance of the model that does not satisfy the property.
Next, {\tool} uses a partial max sat (PMAXSAT) solver to find an instance that does satisfy the property and is as close as possible to the counterexample.
{\tool} then determines the relations and atoms that are different between the cex and sat instance.
Finally, {\tool} analyses these differences to compute suspicion scores for expressions in the original model.

Unlike AlloyFL that relies on unconventional unit tests,  
{\tool} uses well-established and widely-used assertions, 
naturally compatible with the development practices in Alloy.
Also, instead of using mutation testing or statically analyzing the effects of tests, {\tool} relies on counterexamples and satisfying instances generated by constraint solvers, which are the main underlying technology in Alloy.

We evaluated {\tool} on a benchmark consisting of a suite of 
buggy models from AlloyFL~\cite{alloyfl}. The experimental results corroborate that {\tool} is able to consistently rank buggy expressions in the top 1.9\% of the suspicious list. We also evaluated {\tool} on three case studies consisting of larger Alloy models used in the real-world settings (e.g., Alloy model for surgical robots, Java programs and Android permissions), and {\tool} was able to identify the buggy expressions within the top 1\%.
The run time of {\tool} for most the 
models is under 5 seconds (under 1 second for the AlloyFL benchmarks).
The experimental results corroborate that
{\tool} has the potential to facilitate a non-trivial task of formal specification development significantly
and exposes opportunities for researchers to exploit new debugging techniques for Alloy. 

To summarize, this paper makes the following contributions:
\begin{itemize}
\item \textit{Fault localization approach for declarative models}:
We present a novel fault localization approach for declarative models specified in the Alloy language. The insight underlying our approach is that expressions in an Alloy model that likely cause an assertion violation can be identified by analyzing the counterexamples and closely related satisfying instances.

\item \textit{Tool implementation}: We develop a fully automated
technology, dubbed {\tool}, that effectively realizes our
fault localization approach. We make {\tool} publicly available to the research and education~community~\cite{BFA}.

\item \textit{Empirical evaluation}: We evaluate {\tool} in the context of faulty Alloy specifications found in prior work and specifications derived from real-world systems, corroborating {\tool}’s ability to  consistently rank buggy expressions high on the suspicious list, and analyze complex, real-world Alloy models with thousand lines of code.
\end{itemize}
 
The rest of the paper is organized as follows. Section~\ref{Illustration} motivates our research through an illustrative example. Section~\ref{sec:approach} describes the details of our fault localization approach for Alloy models. Section~\ref{sec:eval} presents the implementation and evaluation of the research. The paper concludes with an outline of the related research and future work.

\section{Illustration}
\label{Illustration}
\begin{figure}[t]
\begin{lstlisting}[language=Alloy, escapechar=?, escapeinside={(*@}{@*)}, basicstyle=\footnotesize\ttfamily]
one sig FSM {
  start: set State,
  stop: set State
}
sig State { transition: set State } (*@\label{line:sig}@*)
fact OneStartAndStop {
  // If a start state exists, there is only one of them
  all start1, start2 : FSM.start | start1 = start2
  // If a stop state exists, there is only one of them
  all stop1, stop2 : FSM.stop | stop1 = stop2
  some FSM.stop
}
fact ValidStartAndStop {
  // start state is not a subset of stop state
  FSM.start !in FSM.stop
  // No transition ends at the start state.
  all s : State | FSM.start !in s.transition (*@\label{line:unsat1}@*)
  // Error: should be "<=>" instead of "=>".
  all s: State | s.transition = none => s in FSM.stop (*@\label{line:bug}@*)
}
fact Reachability {
  // All states are reachable from the start state.
  State = FSM.start.*transition
  // The stop state is reachable from any state.
  all s: State | FSM.stop in s.*transition (*@\label{line:sexp1}@*)
}
assert NoStopTransition{
	no FSM.stop.transition
}
check NoStopTransition for 5
\end{lstlisting}
\caption{Buggy FSM model, adapted from AlloyFL~\cite{alloyfl}.}
\label{fig:ex}
\end{figure}

To motivate the research and illustrate our approach, we provide an Alloy specification of a finite state machine (FSM), adapted from AlloyFL benchmarks~\cite{alloyfl}. 
The specification defines two type signatures, i.e., \texttt{State} and \texttt{FSM}, along with their fields (lines 1--5). The specification contains three fact paragraphs, expressing the constraints, detailed below: If a start (or a stop) state exists, there is only one of them (fact \texttt{OneStartAndStop}). The start state is not a subset of the stop state; 
no transition terminates at the start state; and no transition leaves a stop state (fact \texttt{ValidStartAndStop}).
Finally, every state is reachable from the start state, and  the stop state is reachable from any state (fact \texttt{Reachability}).

\begin{figure}
  \centering  
  \begin{subfigure}{0.48\linewidth}
    \includegraphics[width=1\linewidth]{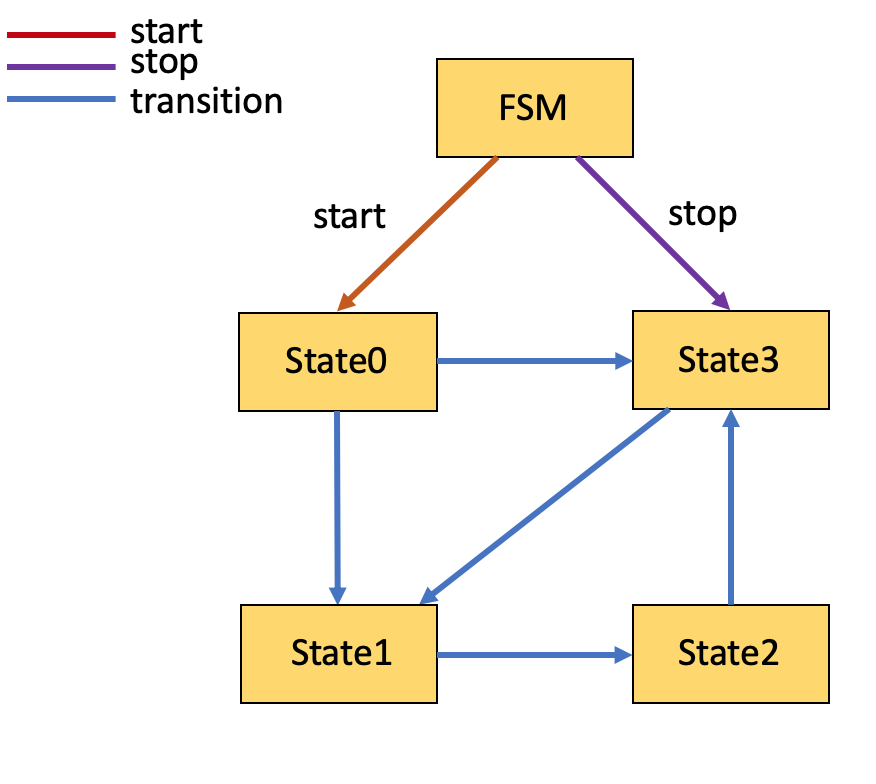}
    \caption{Counterexample}
    \label{fig:cex1}
  \end{subfigure}
  \hfill
  \begin{subfigure}{0.48\linewidth}
    \includegraphics[width=1\linewidth]{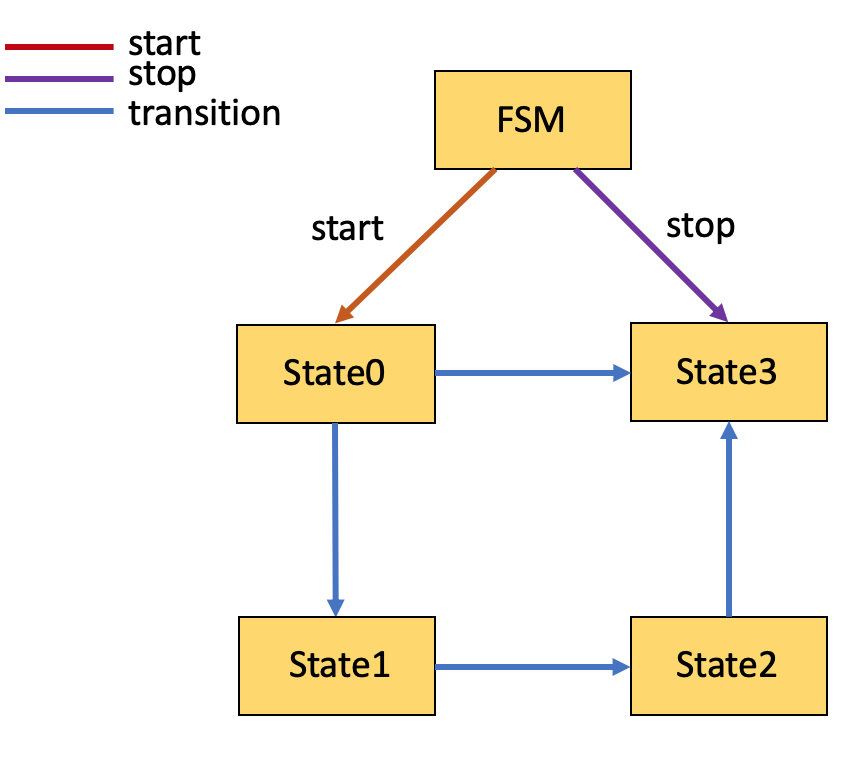}
    \caption{(PMAX) Sat instance}
    \label{fig:expected1}
  \end{subfigure}
  \caption{Instance Pair. Note the similarity between the instances.}
  \label{fig:pair1}
\end{figure}

Each \texttt{assertion} specifies a property that is expected to hold in all instances of the model.
For example, we use the assertion \texttt{NoStopTransition} to check that a stop state behaves as a sink. The Alloy Analyzer disproves this assertion by producing a counterexample, shown in Figure~\ref{fig:cex1}, in which the stop state labeled \texttt{State3} transitions to \texttt{State1}.

Thus, there is a ``bug'' in the model causing the assertion violation.
Indeed, careful analysis of the model and the generated cex reveals that the problem is in the expression on line~\ref{line:bug}: instead of stating that a stop state does not have any transition to any state, the expression states that any state not having a transition to anywhere is a stop state---a subtle logical error that is difficult to realize\footnote{There are two potential fixes for this: (i) reverse the expression to: \texttt{s in FSM.stop => s.transition=none} 
or (ii) replace the implication operator (\texttt{=>}) to logical equivalence (\texttt{<=>}), which technically would strengthen the intended requirement.}.

\begin{table}[h]
  \caption{{\tool}'s results obtained for the model in Figure~\ref{fig:ex}.}
\label{table:model1}
  \small
  \centering
  \begin{tabular}{l  l }
    \hline  
    \hline
    Suspicious Expression & Score \\
     \hline     
     s.transition = none =$>$ s in FSM.stop (=$>$)~\ref{line:bug}  & 1.58 \\
     s.transition = none~\ref{line:bug} & 1.25 \\
     FSM.stop in s.*transtion~\ref{line:sexp1}  &  0.5 \\
     s in FSM.stop~\ref{line:bug}  & 0.5  \\
     \hline
     \hline     
  \end{tabular}
\end{table}

The goal of {\tool} is to identify such buggy expressions automatically.
For this example, within a second, {\tool} identifies four suspicious expressions with the one on line~\ref{line:bug} ranked first. Table~\ref{table:model1} shows the results: expressions or nodes with higher scores are ranked higher.
Moreover, {\tool} suggests that the operator \texttt{=>} is likely the issue in the expression.
Such a level of granularity can significantly help the developer understand and fix the problem.
The results in Section~\ref{sec:eval} show that {\tool} can consistently rank the exact buggy expression within the top 5 suspicious ones and do so under a second.

The key idea underlying our fault localization approach is to analyze the differences between \emph{counterexamples} (instances of the model that do not satisfy the assertion) and instances that do satisfy the assertion to find suspicious expressions in the input model.
{\tool} first checks the assertion \code{NoStopTransition} in the model using the Alloy Analyzer, which returns the cex in Figure~\ref{fig:cex1}.
Next, {\tool} generates a satisfying (sat) instance that is as minimal and similar to the cex as possible.
Their differences promise effective localization of the issue.

\paragraph{Generating SAT instances} To obtain an instance similar to the cex, {\tool} transforms the input model into a logical formula representing \emph{hard constraints} and the information from cex into a formula representing \emph{soft constraints}.
Essentially, {\tool} converts the instance finding problem into a \emph{Partial-Max SAT} problem~\cite{pmaxsat} and then uses the Pardinus~\cite{pardinus} solver to find a solution that satisfies all the hard constraints and as many soft constraints as possible.
Thus, the result is an instance of the model that is similar to the cex but satisfies the assertion.
Figure~\ref{fig:expected1} shows an instance produced by Pardinus, considering the cex shown in Figure~\ref{fig:cex1}.
Notice that this instance is similar to the given cex, except for the edge from \texttt{State3} to \texttt{State1}, which represents the main difference between the two instances.

\paragraph{Finding Suspicious Expressions} {\tool} analyzes the differences between cexs and the sat instances---e.g., here the transition from \texttt{State3} to \texttt{State1}, which only appears in the cex but not the sat instance---to identify Alloy relations causing the issue.
As shown in Table~\ref{fig:textrepr}, that demonstrates the Alloy text representation of the cex in Figure~\ref{fig:cex1}, the \texttt{transition} relation involves the tuple \texttt{State3 -> State1} and the \texttt{stop} relation involves \texttt{State3}.
Thus, {\tool} hypothesizes that two relations of \texttt{transition} and \texttt{stop} may cause the difference in the two models.
Note that while we present one cex and one sat instance in this example for the sake of simplicity, {\tool} supports analyzing multiple pairs of cex and sat instances in tandem. 

\begin{table}[h]
  \caption{Text Representation of the cex in Figure~\ref{fig:cex1}}
  \label{fig:textrepr}
  \begin{tabular}{|c|c|}
    \hline
    Relation  &  Tuples \\ 
    \hline
    FSM       &  FSM0  \\
    \hline
    start     &  FSM0-$>$State0 \\
    \hline
    stop      &  FSM0-$>$State3 \\
    \hline
    State     &  State0, State1, State2, State3 \\
    \hline
    \multirow{2}{4em}{transition} & State0-$>$State1, State0-$>$State3, State1-$>$State2, \\
                             & State2-$>$State3, State3-$>$State1 \\ 
    \hline
  \end{tabular}
\end{table}




Next, {\tool} \emph{slices} the input model to contain only expressions affecting both relations \texttt{transition} and \texttt{FSM.stop}.
This results in two expressions:
  \texttt{all s: State | FSM.stop in s.*transition} (line~\ref{line:sexp1})
  and
  \texttt{all s: State | s.transition = none => s in FSM.stop} (line ~\ref{line:bug}).

At this point, {\tool} could stop and return these two expressions, one of which is the buggy expression on line~\ref{line:bug}.
Indeed, this level of ``statement'' granularity is often used in fault localization techniques, like Tarantula~\cite{tarantula} or Ochiai~\cite{spfl1}.
However, {\tool} aims to achieve a finer-grained granularity level by also considering the \emph{boolean} and \emph{relational} subexpressions, detailed below.

\paragraph{Ranking Boolean Nodes} The expressions on lines~\ref{line:sexp1} and~\ref{line:bug} have four boolean nodes: (a) \texttt{FSM.stop in s.*transition}, (b) \texttt{s.transition = none}, (c) \texttt{s in FSM.stop}, and (d) \texttt {s.transition = none => s in FSM.stop}.
{\tool} instantiates each of these with \texttt{State1} and \texttt{State3}, the values that differentiate the cex and sat instance.
For example, (a) becomes \texttt{FSM.stop in S1.*transition} and \texttt{FSM.stop in S3.*transition}. Next, {\tool} evaluates these instantiations using the cex and sat instance and assigns a higher suspicious score to those with inconsistent evaluation results.
For example, the instantiations \texttt{FSM.stop in S1.*transition} and \texttt{FSM.stop in S3.*transition} of node (a) evaluate to \texttt{true} in both cex and sat instance, so we give (a) the score 0 (i.e., no changes).
We assign score 1 to (b) because \texttt{State3.transition = none} evaluates to \texttt{false} in the cex but \texttt{true} in the sat instance (thus 1 change) and \texttt{State1.transition = none} evaluates to false in both (no change).

Overall, {\tool} obtains the scores 0, 1, 0, 1 for nodes (a), (b), (c), (d), respectively.
Thus, {\tool} determines that nodes (b) \texttt{s.transition = none} and (d) \texttt {s.transition = none => s in FSM.stop} are the two most suspicious boolean subexpressions within the expression on line~\ref{line:bug}.





\paragraph{Ranking Relational Nodes}  While subexpression (d) indeed contains the error, it receives the same score as subexpression (a).
To achieve more accurate results\footnote{While this example has only two expressions with similar scores,  we obtain many expressions with similar scores in more complex and real-world models. Thus, this step is crucial to distinguish the buggy expressions from the rest.}, {\tool} further analyzes the involving relations.
{\tool} instantiates these relations with \texttt{State1} and \texttt{State3}, assesses them in the context of the cex and sat instances, and assigns scores based on the evaluations.
For example, node (d) \texttt{s.transition = none => s in FSM.stop} contains 3 relations:
(1) \texttt{s.transition},
(2) \texttt{s}, and
(3) \texttt{FSM.stop}.
Instantiating these relations with \texttt{State3} and evaluating them using the cex is as follows:
(1) becomes \{\texttt{State1}\},
(2) \{\texttt{State3}\}, and
(3) \{\texttt{State3}\}.
Thus, for the cex, (d) involves both \texttt{\texttt{State1}} and \texttt{\texttt{State3}}, and {\tool} gives it a score of 1.
Next, it evaluates the instantiations using the sat instance as follows:
(1) becomes \{\},
(2) \{\texttt{State3}\}, and
(3) \{\texttt{State3}\}.
(d) does not involve \texttt{State1} and thus has a score 0.
{\tool} assigns (d) the average score of 0.5 (for the instantiations of \texttt{State3}).
Performing a similar computation for the instantiation of \texttt{State1}, we obtain a score of 2/3 for (d) as the evaluation for the cex and sat instances involves both \texttt{State1} and \texttt{State3} (differentiated values) and \texttt{State2} (regular value). 
Thus, (d) has a score of 0.58 as an average of 0.5 and 2/3.

Overall, {\tool} obtains the scores 0.5, 0.25, 0.5,  0.58 for (a), (b), (c), and (d), respectively.
Note that the node (d) is now ranked higher than (a) as desired.

\paragraph{Suspicious Scores}
{\tool} computes the final suspicious score of a node as the sum of the boolean and relational scores of that node, as shown above.
For example, the node (d) \texttt{s.transition = none => s in FSM.stop} in the expression on line~\ref{line:bug} has the highest suspicious score of 1.58.
Table~\ref{table:model1} shows suspicious scores of the expressions in the ranked list returned by {\tool}.

In addition, 
{\tool} analyzes (non-atomic) nodes containing (boolean) connectors and reports connectors that connect subnodes with different scores.
For example, {\tool} suggests that the operator \texttt{=>} is likely responsible for the error in (d) because the two subexpressions \texttt{s.transition = none} and \texttt{s in FSM.stop} have different scores as shown in Table~\ref{table:model1}.
Indeed, in this example, the assertion violation is entirely due to this operator (a potential fix would be strengthening the model to \texttt{<=>} or switching the two subexpressions as Alloy does not have the operator \texttt{<=}).

\section{Approach}
\label{sec:approach}
\SetKwInOut{Input}{input}
\SetKwInOut{Output}{output}
\SetKwData{pairs}{pairs}
\SetKwData{maxpairs}{max\_instance\_pairs}
\SetKwData{cex}{c}
\SetKwData{sin}{s}
\SetKwData{diffs}{diffs}
\SetKwData{diff}{diff}
\SetKwData{nil}{nil}
\SetKwData{diff}{diff}
\SetKwData{expr}{expr}
\SetKwData{sexpr}{sexpr}
\SetKwData{exprs}{exprs}
\SetKwData{sexprs}{sexprs}
\SetKwData{conflicts}{conflicts}
\SetKwData{false}{false}
\SetKwData{rel}{rel}
\SetKwData{relates}{relates}
\SetKwData{infers}{infers}
\SetKwData{name}{name}
\SetKwData{tuples}{tuples}
\SetKwData{tuple}{tuple}
\SetKwData{atom}{atom}
\SetKwData{crel}{crel}
\SetKwData{srel}{srel}
\SetKwData{rank}{rank}
\SetKwData{sus}{suspicious}
\SetKwData{suspexprs}{susp\_exprs}
\SetKwData{keywords}{keywords}
\SetKwData{subs}{subs}
\SetKwData{sub}{sub}
\SetKwData{esubs}{epxr\_subs}
\SetKwData{node}{node}
\SetKwData{score}{score}
\SetKwData{relscore}{relscore}
\SetKwData{results}{results}
\SetKwData{instexpr}{isexpr}
\SetKwData{vals}{vals}
\SetKwData{cvals}{cvals}
\SetKwData{svals}{svals}
\SetKwData{instscore}{instscore}
\SetKwData{pexpr}{pexpr}
\SetKwData{child}{child}

\SetKwFunction{getchildren}{getchildren}
\SetKwFunction{AlloyAnalyzer}{AlloyAnalyzer}
\SetKwFunction{Pardinus}{Pardinus}  
\SetKwFunction{PMaxSolver}{PMaxSolver}
\SetKwFunction{AlloySolver}{AlloySolver}
\SetKwFunction{block}{blockcex}
\SetKwFunction{comparator}{comparator}
\SetKwFunction{gen}{gencex}
\SetKwFunction{unsat}{get\_unsatcore}
\SetKwFunction{unsatanalyzer}{unsat\_analyzer}
\SetKwFunction{diffsanalyzer}{diffs\_analyzer}
\SetKwFunction{getdiffs}{get\_diffs}
\SetKwFunction{getdiff}{get\_diff}
\SetKwFunction{slice}{slice}
\SetKwFunction{eval}{eval}
\SetKwFunction{collect}{collect\_exprs}
\SetKwFunction{infer}{infer}
\SetKwFunction{extract}{extract\_keywords}
\SetKwFunction{sort}{sort}
\SetKwFunction{inst}{instantiate}
\SetKwFunction{isbool}{isbool}
\SetKwFunction{getrelsubs}{get\_rel\_subs}
\SetKwFunction{getsuspexprs}{get\_susp\_exprs}
\SetKwFunction{collectsexprs}{collect\_exprs}
\SetKwFunction{compute}{computescore}
\SetKwFunction{isleaf}{isleaf}

\SetKwProg{Func}{Function}{:}{}

\begin{figure}[t!]
    \centering
    \includegraphics[width=\linewidth]{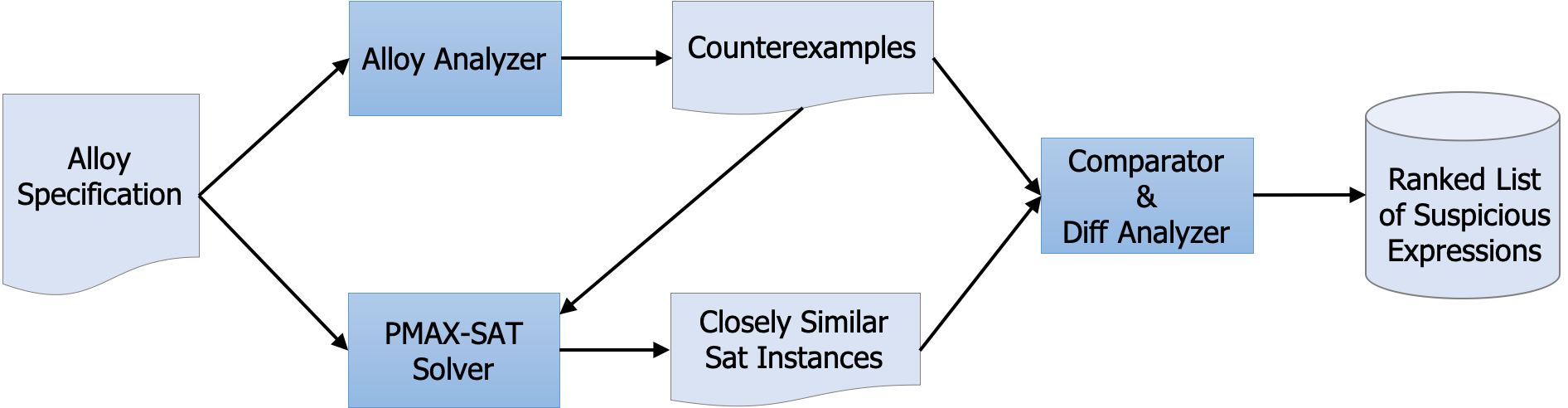}
    \caption{Overview of {\tool}.}
    \label{fig:workflow}
\end{figure}

Figure~\ref{fig:workflow} gives an overview of {\tool}, which takes as input an Alloy model with some violated assertion and returns a ranked list of suspicious expressions contributing to the assertion violation.
The insight guiding our research is that the differences between counterexamples that do not satisfy the assertion and closely related satisfying instances can drive localization of suspicious expressions in the input model.
To achieve this, {\tool} uses the \emph{Alloy analyzer} to find counterexamples showing the violation of the assertion.
It then uses a PMAX-SAT solver to find satisfying instances that are \emph{as close as possible} to the cex's.
Next, {\tool} analyzes the differences between the cex's and satisfying instances to find expressions in the model that likely cause the errors.
Finally, {\tool} computes and returns a ranked list of suspicious expressions.

\subsection{The Alloy Analyzer}

An Alloy specification or \emph{model} consists of three components: (i) Type signatures (\texttt{sig}) define essential \emph{data types} and their \emph{fields} capture relationships among such data types, (ii) \texttt{fact}s, predicates (\texttt{pred}), and assertions (\texttt{assert}) are formulae defining constraints over data types, and (iii) \texttt{run} and \texttt{check} are commands to invoke the Alloy Analyzer. The \texttt{check} command is used to find counterexamples  violating some asserted property, and \texttt{run} finds 
satisfying model instances (\emph{sat instances}).
For a model $M$ and a property $p$, a cex is an instance of $M$ that satisfies $M \wedge \neg p$, and a sat instance is one that satisfies $M \wedge p$.
The specification in Figure~\ref{fig:ex} defines two signatures (\texttt{FSM, State}), three fields (\texttt{start, stop, transition}), three facts (\texttt{OneStartStop, ValidStart, ValidStop}) and one assertion (\texttt{NoStopTransition}).



Analysis of specifications written in Alloy is entirely automated, yet bounded up to user-specified scopes on the size of type signatures.
More precisely, to check that  $p$ is satisfied by \emph{all} instances of $M$ (i.e., $p$ is valid) up to a certain scope, the Alloy developer encodes $p$ as an \emph{assertion} and uses the \texttt{check} command to validate the assertion, i.e., showing that no cex exists within the specified scope (a cex is an instance $I$ such that $I \vDash M \land \neg p$).
To check that $p$ is satisfied by \emph{some} instances of $M$, the Alloy developer encodes $p$ as a predicate and uses the \texttt{run} command to analyze the predicate, i.e., searching for a sat instance $I$ such that $I \vDash M \land p$.
In our running example, the \texttt{check} command examines the \texttt{NoStopTransition} assertion and returns a cex in Figure~\ref{fig:cex1}.

Internally, Alloy converts these tasks of searching for instances into boolean formulae and uses a SAT solver to check the satisfiability of the formulae. Each value of each relation is translated to a distinct variable in the boolean formula. For example, given a scope of 5 in the FSM model in Figure~\ref{fig:ex}, the relation \texttt{State} contains 5 values and is translated to 5 distinct variables in the boolean formula, and the \texttt{transition} is translated to 25 values representing 25 values of combinations of $\|State\| \times \|State\|$. An instance is an assignment for all variables that makes the formula True. For example, \emph{cex} in Figure~\ref{fig:cex1} is an assignment where all variables corresponding to values in Table~\ref{table:model1} are assigned True and others are False.
Finally, Alloy translates the result from the SAT solver, e.g., an assignment that makes the boolean formula True, back to an instance of $M$.

\subsection{The {\tool} Algorithm}
\label{sec:alg}

\begin{algorithm}[t!]
  \footnotesize
  \DontPrintSemicolon
  
  \SetAlgoLined
  \Input{Alloy model $M$, property $p$ not satisfied by $M$}
  \Output{Ranked list of suspicious expressions in $M$}
  $\AlloySolver \leftarrow \AlloyAnalyzer(M, p)$\;
  $\pairs \leftarrow \emptyset$\;
  \While{$|\pairs| <  \maxpairs$}{
    $\cex \leftarrow \AlloySolver.\gen()$\;
    $\AlloySolver.\block(\cex)$\;
    $\sin \leftarrow \PMaxSolver(M, \cex)$\;
    \uIf{$\sin = \nil$}{
       $U \leftarrow \AlloySolver.\unsat()$\;
       \Return $\unsatanalyzer(M, U, \cex)$\;
     } 
      
    $\pairs \leftarrow \pairs \cup (\cex, \sin)$\;
  }
  $\diffs \leftarrow \comparator(\pairs)$\;
  \Return $\diffsanalyzer(\diffs)$\;
  \caption{{\tool} fault localization process}
  \label{algo:overview}
\end{algorithm}

Algorithm~\ref{algo:overview} shows the algorithm of {\tool}, which takes as input an Alloy model $M$ and a property $p$ that is not satisfied by $M$ (as an assertion violation) and returns a ranked list of expressions that likely contribute to the assertion violation.
{\tool} first uses the Alloy Analyzer and the Pardinus PMAX-SAT solver to generate pairs of cex and closely similar sat instances. 
{\tool} then analyzes the differences between the cex and sat instances to locate the error.
If {\tool} cannot generate any sat instance, {\tool} inspects the \emph{unsat core} returned by the Alloy Analyzer to locate the error.

\subsubsection{Generating Instances}
\label{sec:geninsts}
\begin{figure}[h]
  \centering
  \begin{subfigure}{0.47\linewidth}
    \includegraphics[width=1\linewidth]{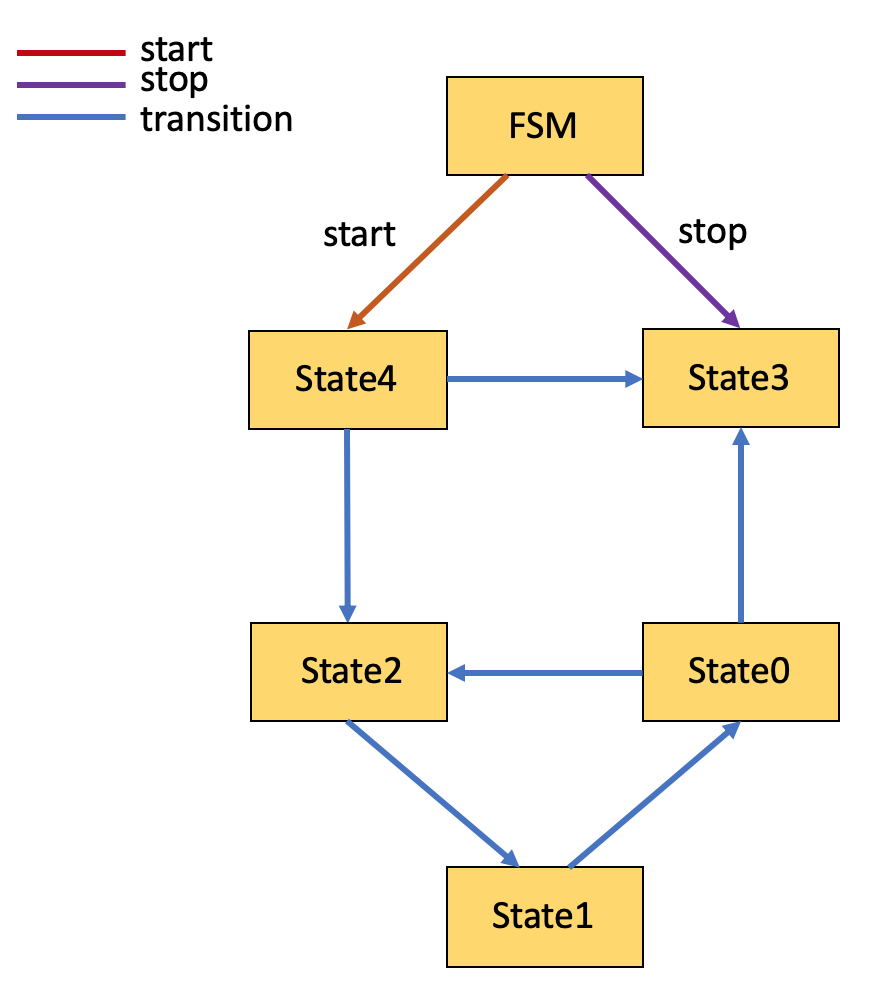}
  \end{subfigure}%
  \hfill
  \begin{subfigure}{0.47\linewidth}
    \includegraphics[width=1\linewidth]{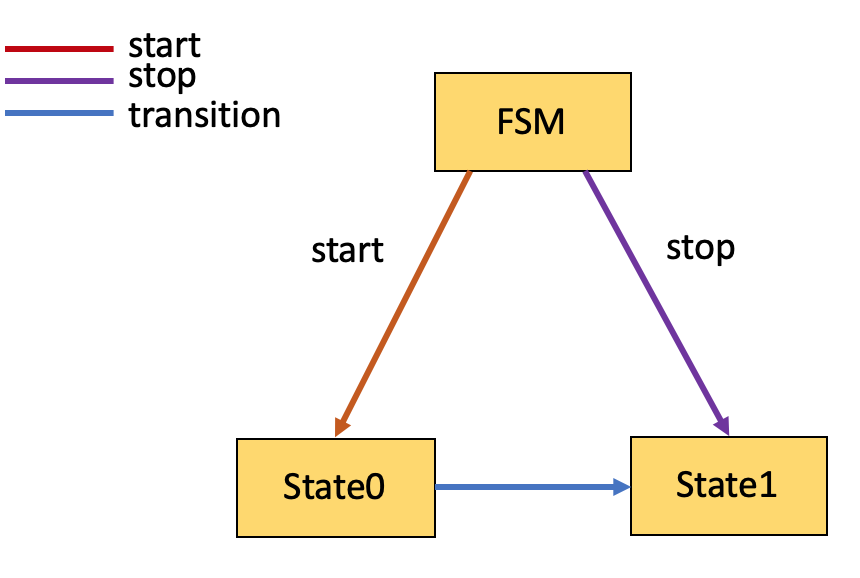}
  \end{subfigure}
  \caption{Model instances generated by the Alloy Analyzer.}
  \label{fig:sat_inst_alloy}
\end{figure}

To understand why $M$ does not satisfy $p$, {\tool} obtains differences between cexs and relevant  sat instances.
These differences can lead to the cause of the error.
One option is to use the Alloy Analyzer to generate a sat instance directly (e.g., by checking a \emph{predicate} consisting of $p$).
However, such an instance generated by Alloy is often predominantly different from the cex, and thus does not help identify the main difference.
For example, the cex, shown in Figure~\ref{fig:cex1}, that violates the assertion \texttt{NoStopTransition} is quite different from the two Alloy-generated satisfying instances, shown in Figure~\ref{fig:sat_inst_alloy}.




To generate a sat instance closely similar to the cex, we reduce the problem to a PMAX-SAT (partial maximum satisfiability) problem.

\begin{definition}[Finding a Similar Sat Instance from a Cex]
Given a set of \emph{hard} clauses, collectively specified by model $M$ and property $p$, and a set of \emph{soft} clauses, specified by a counterexample $cex$, find a solution that satisfies \emph{all} hard clauses and satisfies \emph{as many soft clauses as possible}. 
\end{definition}

More specifically, 
the hard clauses are generated by constraints in the Alloy model, 
and the soft clauses are the assignment represented by \emph{cex} where all presenting variables are True and other variables are False. Because the relations and scope of the Alloy model stay the same, the variables in the transformed boolean formula would also remain the same, and just the values assigned to them would differ between various model instances. Thus, this encoding can apply to general instances regardless of their structures.

{\tool} then uses an existing PMAX-SAT solver (Pardinus) to find an instance that has the property $p$ and is as similar to the cex as possible\footnote{Based on our experiment, the first solution returned by the PMAX-SAT solver is similar enough for {\tool} to locate bugs.}.
For example, the instance in Figure~\ref{fig:expected1} generated by Pardinus is similar to the cex in Figure~\ref{fig:cex1}, but has an extra edge from  \texttt{State3} to \texttt{State1}.
The idea is that such (minimal) differences can help {\tool} identify the error.







\subsubsection{Comparator}
\label{sec:comparator}

{\tool} compares the generated cex's and sat instances to obtain their differences, which involve atoms, tuples, and relations.
First, it obtains tuples and their atoms that are different between the cex and sat instance, e.g., in Figure~\ref{fig:pair1}, the tuple \texttt{State3->State1}, which has the atoms \texttt{State1} and \texttt{State3}, is in the cex but not in the satisfying instance. Next, it obtains relations with different tuples between the cex and sat instance, e.g., the \texttt{transition} relation involves the tuple \texttt{State3->State1} in the cex but not in the sat instance. Third, it obtains relations that can be inferred from the tuples and atoms derived in the previous steps, e.g., the relation \texttt{FSM.stop} involves tuples having the \texttt{State3} atom. 

In summary, for the pair of cex and sat instance in Figure~\ref{fig:pair1}, {\tool} obtains the suspicious relations \texttt{transitions} and \texttt{stop} and the atoms \texttt{State1, State3}. {\tool} applies these comparison steps for all pairs of instances and cex's and uses the common results.

\subsubsection{Diff Analyzer}
After obtaining the differences consisting of relations and atoms between cex's and sat instances, {\tool} analyzes them to obtain a ranked list of expressions based on their suspicious levels.
{\tool} assigns higher suspicious scores to expressions whose evaluations depend on these differences (and lower scores to those not depending on these differences).

Algorithm~\ref{algo:diff} shows the Diff Analyzer algorithm, which takes as input a model $M$, the differences $\diffs$ obtained in Sect~\ref{sec:comparator}, and pairs of cex and sat instances obtained in Sect~\ref{sec:geninsts}, and outputs a ranked list of suspicious expressions in $M$.
It first identifies expressions in $M$ that involve relations in $\diffs$.
These expressions are likely related to the difference between cex and sat instance.
For example, consider the model in Figure~\ref{fig:ex}. {\tool} identifies two expressions: \\ \texttt{all s: State | FSM.stop in s.*transition} on line~\ref{line:sexp1} and \texttt{all s: State | s.transition = none => s in FSM.stop} on line~\ref{line:bug}, as they involve the relations \texttt{transition} and \texttt{stop} in $\diffs$.

\begin{algorithm}[t]
    \footnotesize
    \DontPrintSemicolon
    \SetAlgoLined
    \Input{Alloy model $M$, differences $\diffs$, pairs of cex's and sat instances $\pairs$ }
    \Output{Ranked list of suspicious expressions in $M$}
    $\exprs \leftarrow \getsuspexprs(M, \diffs)$\;
    $\results \leftarrow \{\}$\;
    \ForEach{$\expr \in \exprs$}{
      $\compute{\expr, \results}$\;
    }
    \Return $\sort{\results}$\;

    \BlankLine

    \Func{\compute{$\expr, \results$}}{
      $\score = 0$\;
      \uIf{\isleaf($\expr$)}{
        $\instexpr \leftarrow \inst(\expr,\diff)$\;
        \ForEach{$(\cex,\sin) \in \pairs$}{
          $\cvals \leftarrow \eval(\cex,\instexpr)$\;
          $\svals \leftarrow \eval(\sin,\instexpr)$\;
          $\instscore \leftarrow 0$\;
          \uIf{$\diff \subset \cvals$}{
            $\instscore \leftarrow \instscore + \frac{|\diff|}{|\cvals|}$\;
          }
          \uIf{$\diff \subset \svals$}{
            $\instscore \leftarrow \instscore + \frac{|\diff|}{|\svals|}$\;
          }
          $\score \leftarrow \score + \frac{\instscore}{2}$)\;
        }
        $\score \leftarrow \frac{\score}{|\pairs|}$\;
      }
      \Else{
        \uIf{$\isbool(\expr)$}{
          $\instexpr \leftarrow \inst(\expr,\diff)$\;
          \ForEach{$(\cex,\sin) \in \pairs$}{
            \uIf{$\eval(\cex,\instexpr) \neq \eval(\sin,\instexpr)$}{
              $\score \leftarrow \score + 1$\;
            }
          }
        }
        \ForEach{$\child \in \getchildren(\expr)$}{
          $\score \leftarrow \score + \compute(\child, \results)$\;
        }
      }
      $\results \leftarrow \results \cup (\expr, \score)$\;
      \Return $\score$\;
    }
  \caption{Diff Analyzer}
  \label{algo:diff}
\end{algorithm}

{\tool} then recursively computes the suspicious score for each collected expression $e$, represented as an AST tree.
If $e$ is a leaf (e.g., a relational expression), {\tool} instantiates $e$ with atoms from $\diffs$.
{\tool} then evaluates the instantiated expression for each pair of cex and sat instance.
If the evaluated result for an instantiated expression contains \textit{all} atoms involved in $\diffs$, {\tool} computes the score as ``\textit{size of $\diffs$ / size of evaluated results};'' otherwise, the score is 0. For a pair, the score is then the average score of cex and sat instance. At last, the score of $e$ is the average among all pairs.
Essentially, a higher suspicious score is assigned to a relational subexpression whose evaluation involves many atoms in $\diffs$.

If $e$ is not a leaf node, $e$'s score is the sum of boolean and relational scores. 
If $e$ is a boolean expression (i.e., an expression that returns \texttt{true} or \texttt{false}), we instantiate $e$ with atoms from $\diffs$ and evaluate it on each cex and sat instance pair.
If it evaluates to different results between the cex and sat instance (e.g., one is \texttt{true} and the other is \texttt{false}), 
{\tool} increases $e$'s score by 1.
Thus, a higher boolean score is assigned to the expressions whose evaluation does not match between pairs of the cex and sat instances.
Then $e$'s relational score is calculated as the sum of $e$'s children. 
The final score assigned to each expression is the sum of the $e$'s boolean scores and the relational scores of $e$'s children.
In the end, {\tool} returns all expression ranked by their suspicious scores.
To make the idea concrete, consider the expression
\texttt{s.transition = none} in Figure~\ref{fig:ex}. For the cex and sat instance pair in Figure~\ref{fig:pair1}, $\diffs$ contain two atoms \texttt{State1} and \texttt{State3}.
{\tool} first instantiates the expression under analysis with the atoms mentioned above into two concrete expressions: (1) \texttt{State1.transition = none} and (2) \texttt{State3.transition = none}. The concrete expression (1) evaluates to \texttt{false} in both cex and sat instance, while the concrete expression (2) evaluates to
\texttt{true} in cex and \texttt{false} in the sat instance.
Thus, the boolean score for the expression under analysis is 1 as the aggregation of the values obtained for the concrete expressions (1) and (2).

{\tool} then computes the relational score for the expression under analysis as the sum of the relational scores for its children: \texttt{s.transition} and \texttt{none}, both of which are leaves. 
To compute the score for \texttt{s.transtion}, it is instantiated to \texttt{State1.transition} and \texttt{State3.transtion}. \texttt{State1.transition} evaluates to \texttt{State2} in both cex and sat instance. Thus, it gets a score of 0.
For \texttt{State3.transition}, in cex, it evaluates to \texttt{State1} and gets a score of 1 as the size of different values \{\texttt{State3, State1}\} divided by the size of the instantiated values \{\texttt{State3}\} and the evaluated values \{\texttt{State1}\}. In sat instance, it evaluates to an empty set and gets a score of 0.
Overall, \texttt{s.transition} gets a relational score of 0.25 as the average of all its instantiated expressions: \texttt{State1.transition} (0) and \texttt{State3.transition} (0.5). Finally, the overall score of 1.25 is assigned to the expression \texttt{s.transition = none} as the aggregation of its boolean and relational scores. 

\subsubsection{UNSAT Core Analyzer}
\label{sec:unsat}
It is possible that we can only generate cex's, but no sat instances, indicating that some constraints in the model have conflicts with the property we want to check.
To identify these constraints, {\tool} inspects the \emph{unsat core} returned by the Alloy Analyzer.
The unsat core explains why a set of constraints cannot be satisfied by giving a minimal subset of conflicting constraints.
Those conflicting constraints can help {\tool} identify suspicious expressions.

\begin{algorithm}[t]
  \footnotesize
  \DontPrintSemicolon
  \SetAlgoLined  
  \Input{Alloy model $M$, unsat core $U$, counterexample $c$}
  \Output{a set of expressions in $M$}
  $M' \leftarrow \slice(M, U)$\;
  $\sin \leftarrow \PMaxSolver(M', \cex)$\;
  $\diffs \leftarrow \comparator(\{(\cex, \sin)\})$\;
  $\exprs \leftarrow \collect(U)$\;
  $\conflicts \leftarrow \emptyset$\;
  \ForEach{$\expr \in \exprs$}{
    \ForEach{$\diff \in \diffs$}{
      \uIf{$\eval(\expr, \diff, M') = \false$}{
        $\conflicts \leftarrow \conflicts \cup \expr$\;
      }
    }
  }
  \Return $\conflicts$\;
  \caption{UNSAT Analyzer}
  \label{algo:unsatanalyzer}
\end{algorithm}

Algorithm~\ref{algo:unsatanalyzer} outlines the process underlying our UNSAT core analyzer, 
which takes as input a model $M$, an unsat core $U$, and a cex $c$ showing $M$ does not satisfy a property $p$, and outputs a list of expressions in $M$ conflicting with $p$.
Recall that these values, $M$, $U$, and $c$, are earlier inferred by {\tool} as outlined in Algorithm~\ref{algo:overview}.

{\tool} starts by producing a sliced model $M'$, in which  all expressions in the unsat core are omitted from the original model $M$. Removing these conflicting expressions would allow us to obtain sat instances from the new model $M'$ to compare with the cex.
{\tool} now generates a minimal sat instance from $M'$ and compares it with the input cex to obtain the differences between the cex and the sat instance as shown in Section~\ref{sec:comparator}.
Then, {\tool} attempts to identify which of the removed expressions really conflict with $p$ by evaluating them on obtained differences.
The idea is that if an expression evaluates to true, then adding that expression back to the model would still allow the sat instance to be generated, i.e., that expression is not conflicting with $p$.
Thus, expressions that evaluate to \texttt{false} are ones conflicting with $p$ and are returned as suspicious expressions.
Note that we assign similar scores to these resulting expressions because they all contribute to the unsatisfiability of the original model and the intended property.


For example, if we change line~\ref{line:unsat1} in the model shown in Figure~\ref{fig:ex} to \texttt{all s: State | s.transition !in FSM.start}, Alloy would find counterexamples such as the one in Figure~\ref{fig:cex1}, but fail to generate any sat instances. This is because the modified line forces all states to have some transitions, which conflicts with the constraint requiring no transition for stop states. 

From the unsat core, 
{\tool} identifies four expressions in the model: (a) \texttt{all start1, start2 : FSM.start | start1 = start2}, (b) \texttt{some FSM.stop}, (c) \texttt{FSM.start !in FSM.stop} and (d) \texttt{all s: State | s.transition !in FSM.start}. After removing these four expressions from the model, {\tool} can now generate the same sat instance in Figure~\ref{fig:expected1}.
As before, the main difference between the cex and sat instance involves two values:  \texttt{State1} and \texttt{State3}.
Then, {\tool} evaluates each expression using these values. Expressions (a), (b), and (c) are evaluated to \texttt{true} for both values, while expression (d) evaluates to \texttt{false} for \texttt{State3}. Thus, {\tool} correctly identifies (d) as the suspicious expression.

\section{Evaluation}
\label{sec:eval}

{\tool} is implemented in Java 8 and uses Alloy 4.2.
We extend the backend KodKod solver~\cite{kodkod} in Alloy to use the Pardinus solver~\cite{pardinus} to obtain sat instances similar to counterexamples.
We also modify the AST expression representation in Alloy to collect and assign suspicious scores to boolean and relational subexpressions. 

Our evaluation addresses the following research questions:
\begin{itemize}
\item \textbf{RQ1}: Can {\tool} effectively find suspicious expressions?
\item \textbf{RQ2}: How does {\tool} scale to large, complex models?
\item \textbf{RQ3}: How does {\tool} compare to AlloyFL? 
\end{itemize}

All experiments described below were performed on a Macbook with 2.2 GHZ i7 CPU and 16 GB of RAM.

\subsection{RQ1: Effectiveness}
\label{sec:arepair}

To investigate the effectiveness of {\tool}, we use the benchmark models from AlloyFL~\cite{alloyfl}.
Table~\ref{table:res1} shows 152 buggy models collected from 12 Alloy models in AlloyFL. 
These are real faults collected from Alloy release 4.1, Amalgam~\cite{whyandwhynot}, and Alloy homework solutions from graduate students.
Briefly, these models are \emph{addr} (address book) and \emph{farmer} (farmer cross-river puzzle) from Alloy; \emph{bempl} (bad employer), \emph{grade} (grade book) and \emph{other} (access-control specifications) are from Amalgam~\cite{whyandwhynot}; and \emph{arr} (array), \emph{bst} (balanced search tree), \emph{ctree} (colored tree), \emph{cd} (class diagram), \emph{dll} (doubly linked list), \emph{fsm} (finite state machine), and \emph{ssl} (sorted singly linked list) are homework from AlloyFL.

For models with assertions (e.g., from Amalgam~\cite{whyandwhynot}), we use those assertions for the experiments.
For models that do not have assertions (e.g., homework assignments), we manually create assertions and expected predicates by examining the correct versions or suggested fixes (provided by~\cite{alloyfl}).
Moreover, from the correct models or suggested fixes, we know which expressions contain errors and therefore use them as \emph{ground truths} to compare against {\tool}'s results.
{\tool} deals with models containing multiple violated assertions by analyzing them separately and returning a ranked list for each assertion.
For illustration purposes, we simulate this by simply splitting models with separate violations into separate models (e.g., \texttt{bst2} contains two assertion violations and thus are split into two models \texttt{bst2, bst2\_1}). Finally, {\tool} is highly automatic and has just one user-configurable option: the number of pairs of cex and satisfying instances (which by default is set to 5 based on our experiences).
\begin{table*}[]
  \centering
  \caption{Results of {\tool} on 152 Alloy models.  Results are sorted based on ranking accuracy. Times are in seconds.}
  \label{table:res1}
  \begin{tabular}{c|ccc@{\;\;}c@{\;\;}c||c|ccc@{\;\;}c@{\;\;}c||c|ccc@{\;\;}c@{\;\;}c}
    \textbf{model}    & \textbf{loc} & \textbf{total} & \textbf{sliced} & \textbf{rank} & \textbf{time} & \textbf{model}    & \textbf{loc} &  \textbf{total} & \textbf{sliced} & \textbf{rank} & \textbf{time} & \textbf{model} & \textbf{loc} & \textbf{total} & \textbf{sliced} & \textbf{rank} & \textbf{time} \\
    \midrule
    \cellcolor[gray]{.75}\textbf{top 1 (91)} & \cellcolor[gray]{.75}\textbf{41} & \cellcolor[gray]{.75}\textbf{120} & \cellcolor[gray]{.75}\textbf{95} & \cellcolor[gray]{.75}\textbf{1} & \cellcolor[gray]{.75}\textbf{0.2} & ssl10                 & 43          & 155          & 110         & 1            & 0.1          & dll20\_2                      & 36          & 88           & 47          & 2             & 0.0          \\ \cline{1-6}
    addr                & 21          & 74           & 10          & 1          & 0.6          & ssl12                 & 40          & 157          & 114         & 1            & 0.1          & fsm6                          & 29          & 98           & 17          & 2             & 0.0          \\
    arr3                & 24          & 48           & 9           & 1          & 0.1          & ssl14                 & 44          & 158          & 149         & 1            & 0.5          & fsm9\_2                       & 29          & 90           & 18          & 2             & 0.1          \\
    arr4                & 24          & 64           & 61          & 1          & 0.2          & ssl14\_1              & 44          & 158          & 149         & 1            & 0.5          & ssl11                         & 42          & 177          & 127         & 2             & 0.1          \\
    arr5                & 24          & 62           & 59          & 1          & 0.2          & ssl14\_2              & 43          & 153          & 120         & 1            & 0.0          & bst8                          & 59          & 134          & 57          & 3             & 0.3          \\
    arr6                & 25          & 56           & 30          & 1          & 0.3          & ssl14\_3              & 44          & 153          & 108         & 1            & 0.1          & bst8\_1                       & 59          & 134          & 57          & 3             & 0.2          \\
    arr7                & 25          & 63           & 50          & 1          & 0.1          & ssl17                 & 41          & 152          & 119         & 1            & 0.0          & bst22\_1                      & 49          & 199          & 124         & 3             & 0.1          \\
    bst2                & 56          & 134          & 56          & 1          & 0.4          & ssl17\_1              & 42          & 152          & 106         & 1            & 0.1          & dll1\_1                       & 38          & 86           & 57          & 3             & 0.1          \\
    bst2\_1             & 56          & 134          & 95          & 1          & 0.3          & ssl18\_1              & 40          & 160          & 118         & 1            & 0.1          & dll18\_2                      & 36          & 107          & 71          & 3             & 0.6          \\
    bst3\_2             & 55          & 141          & 68          & 1          & 0.2          & ssl18\_2              & 49          & 160          & 85          & 1            & 0.3          & fsm4                          & 31          & 141          & 39          & 3             & 0.0          \\
    cd1                 & 27          & 44           & 33          & 1          & 0.0          & ssl19                 & 40          & 169          & 141         & 1            & 0.1          & fsm5\_2                       & 29          & 69           & 17          & 3             & 0.0          \\
    cd1\_1              & 27          & 44           & 31          & 1          & 0.0          & \textit{arr1}         & 24          & 45           & 31          & 1            & 0.1          & ssl2\_1                       & 44          & 156          & 72          & 3             & 0.1          \\
    cd2                 & 27          & 35           & 25          & 1          & 0.0          & \textit{arr2}         & 25          & 60           & 47          & 1            & 0.2          & ssl13                         & 43          & 174          & 123         & 3             & 0.1          \\
    cd3                 & 26          & 43           & 32          & 1          & 0.0          & \textit{arr10}        & 24          & 59           & 45          & 1            & 0.0          & ssl18                         & 40          & 162          & 131         & 3             & 0.0          \\
    cd3\_1              & 26          & 46           & 31          & 1          & 0.0          & \textit{bst1}         & 51          & 171          & 155         & 1            & 0.2          & arr8                          & 25          & 80           & 15          & 4             & 0.5          \\
    dll1                & 37          & 77           & 63          & 1          & 0.1          & \textit{bst4\_1}      & 52          & 163          & 147         & 1            & 0.1          & bst2\_2                       & 47          & 147          & 92          & 4             & 0.1          \\
    dll2                & 42          & 77           & 63          & 1          & 0.1          & \textit{bst5}         & 52          & 184          & 168         & 1            & 0.1          & bst3                          & 57          & 137          & 97          & 4             & 0.1          \\
    dll3                & 37          & 80           & 59          & 1          & 0.0          & \textit{bst7}         & 52          & 159          & 143         & 1            & 0.1          & fsm2                          & 29          & 70           & 14          & 4             & 0.0          \\
    dll3\_1             & 37          & 75           & 49          & 1          & 0.1          & \textit{bst8\_2}      & 54          & 156          & 140         & 1            & 0.1          & fsm8                          & 29          & 71           & 14          & 4             & 0.1          \\
    dll4                & 37          & 81           & 67          & 1          & 0.1          & \textit{bst9}         & 55          & 185          & 169         & 1            & 0.1          & arr11                         & 24          & 83           & 41          & 5             & 0.1          \\
    dll5\_1             & 39          & 102          & 80          & 1          & 0.1          & \textit{bst10}        & 47          & 157          & 140         & 1            & 0.6          & bst10\_3                      & 55          & 162          & 75          & 5             & 0.3          \\
    dll6                & 36          & 113          & 94          & 1          & 0.1          & \textit{bst10\_2}     & 52          & 172          & 156         & 1            & 0.1          & fsm9\_1                       & 29          & 91           & 18          & 5             & 0.1          \\
    dll7\_1             & 36          & 73           & 59          & 1          & 0.1          & \textit{bst11\_1}     & 60          & 214          & 198         & 1            & 0.1          & ssl15                         & 41          & 161          & 106         & 5             & 0.1          \\ \cline{13-18} 
    dll8                & 36          & 96           & 76          & 1          & 0.1          & \textit{bst13}        & 53          & 200          & 184         & 1            & 0.1          & \cellcolor[gray]{.75}\textbf{top 6-10 (10)}        & \cellcolor[gray]{.75}\textbf{51} & \cellcolor[gray]{.75}\textbf{155} & \cellcolor[gray]{.75}\textbf{73} & \cellcolor[gray]{.75}\textbf{7.6}  & \cellcolor[gray]{.75}\textbf{0.2} \\ \cline{13-18} 
    dll9                & 38          & 100          & 91          & 1          & 0.1          & \textit{bst14}        & 59          & 202          & 186         & 1            & 0.1          & bst3\_1                       & 57          & 137          & 58          & 6             & 0.2          \\
    dll11               & 36          & 87           & 68          & 1          & 0.1          & \textit{bst15}        & 53          & 197          & 181         & 1            & 0.1          & bst20\_1                      & 55          & 152          & 61          & 6             & 0.2          \\
    dll12               & 36          & 77           & 63          & 1          & 0.1          & \textit{bst17\_1}     & 52          & 201          & 185         & 1            & 0.1          & fsm7                          & 29          & 64           & 14          & 6             & 0.1          \\
    dll13               & 36          & 60           & 51          & 1          & 0.0          & \textit{bst18\_1}     & 56          & 204          & 188         & 1            & 0.1          & ssl19\_1                      & 50          & 175          & 106         & 7             & 0.6          \\
    dll14\_1            & 37          & 85           & 71          & 1          & 0.1          & \textit{bst20}        & 52          & 169          & 153         & 1            & 0.1          & bst6                          & 51          & 140          & 61          & 8             & 0.2          \\
    dll15               & 40          & 126          & 107         & 1          & 0.1          & \textit{bst21}        & 52          & 182          & 166         & 1            & 0.2          & bst12\_1                      & 56          & 164          & 68          & 8             & 0.2          \\
    dll16               & 36          & 82           & 68          & 1          & 0.1          & \textit{bst22}        & 50          & 213          & 158         & 1            & 0.6          & bst19\_2                      & 55          & 154          & 86          & 8             & 0.2          \\
    dll17\_1            & 36          & 77           & 63          & 1          & 0.1          & \textit{dll7}         & 38          & 90           & 79          & 1            & 0.1          & ssl19\_2                      & 44          & 174          & 81          & 8             & 0.3          \\
    dll18               & 36          & 102          & 84          & 1          & 0.0          & \textit{dll10}        & 40          & 91           & 85          & 1            & 0.2          & bst17\_2                      & 55          & 177          & 79          & 9             & 0.2          \\
    dll18\_1            & 36          & 101          & 67          & 1          & 0.1          & \textit{dll14}        & 39          & 102          & 91          & 1            & 0.1          & bst22\_2                      & 54          & 209          & 118         & 10            & 0.1          \\ \cline{13-18} 
    dll20               & 36          & 90           & 64          & 1          & 0.0          & \textit{dll17}        & 37          & 89           & 83          & 1            & 0.1          & \cellcolor[gray]{.75}\textbf{\textgreater{}10 (6)} & \cellcolor[gray]{.75}\textbf{51} & \cellcolor[gray]{.75}\textbf{172} & \cellcolor[gray]{.75}\textbf{80} & \cellcolor[gray]{.75}\textbf{12.7} & \cellcolor[gray]{.75}\textbf{0.3} \\ \cline{13-18} 
    farmer              & 99          & 124          & 30          & 1          & 1.6          & \textit{fsm1}         & 31          & 90           & 71          & 1            & 0.0          & bst4\_2                       & 51          & 154          & 61          & 11            & 0.3          \\
    fsm1\_1             & 30          & 90           & 19          & 1          & 0.0          & \textit{fsm3}         & 60          & 67           & 56          & 1            & 0.7          & bst16                         & 64          & 181          & 79          & 12            & 0.3          \\
    fsm7\_1             & 29          & 59           & 14          & 1          & 0.0          & \textit{fsm9}         & 30          & 91           & 79          & 1            & 0.5          & bst17                         & 48          & 186          & 113         & 12            & 0.2          \\
    fsm9\_4             & 30          & 91           & 78          & 1          & 0.1          & \textit{fsm9\_3}      & 32          & 91           & 78          & 1            & 0.1          & ssl12\_1                      & 44          & 161          & 78          & 12            & 0.4          \\ \cline{7-12}
    grade               & 33          & 22           & 7           & 1          & 0.0          & \cellcolor[gray]{.75}\textbf{top 2-5 (35)} & \cellcolor[gray]{.75}\textbf{40} & \cellcolor[gray]{.75}\textbf{120} & \cellcolor[gray]{.75}\textbf{66} & \cellcolor[gray]{.75}\textbf{2.9} & \cellcolor[gray]{.75}\textbf{0.2} & ssl9                          & 44          & 153          & 72          & 13            & 0.1          \\ \cline{7-12}
    ssl1                & 40          & 168          & 126         & 1          & 0.1          & arr7\_1               & 24          & 46           & 9           & 2            & 0.9          & bst22\_3                      & 57          & 199          & 74          & 16            & 0.5          \\ \cline{13-18} 
    ssl2                & 40          & 150          & 122         & 1          & 0.1          & arr9                  & 27          & 83           & 43          & 2            & 0.2          & \cellcolor[gray]{.75}\textbf{fail (10)}            & \cellcolor[gray]{.75}\textbf{42} & \cellcolor[gray]{.75}\textbf{104} & \cellcolor[gray]{.75}\textbf{71} & \cellcolor[gray]{.75}\textbf{-}    & \cellcolor[gray]{.75}\textbf{0.1} \\ \cline{13-18} 
    ssl3                & 45          & 188          & 135         & 1          & 0.0          & bst2\_3               & 52          & 133          & 68          & 2            & 0.1          & bst4                          & 46          & 145          & 95          & -             & 0.1          \\
    ssl3\_1             & 44          & 184          & 121         & 1          & 0.2          & bst12                 & 47          & 146          & 94          & 2            & 0.2          & bst11                         & 54          & 196          & 135         & -             & 0.2          \\
    ssl4                & 40          & 146          & 118         & 1          & 0.1          & bst18                 & 51          & 187          & 115         & 2            & 1.4          & bempl                         & 51          & 14           & 7           & -             & 0.0          \\
    ssl5                & 42          & 188          & 160         & 1          & 0.6          & bst19                 & 47          & 158          & 112         & 2            & 0.1          & ctree                         & 30          & 49           & 5           & -             & 0.0          \\
    ssl6                & 42          & 157          & 148         & 1          & 0.7          & bst19\_1              & 52          & 175          & 112         & 2            & 0.1          & dll3\_2                       & 36          & 75           & 67          & -             & 0.0          \\
    ssl6\_1             & 42          & 157          & 148         & 1          & 0.5          & dll2\_1               & 42          & 82           & 57          & 2            & 0.0          & other                         & 34          & 26           & 19          & -             & 0.0          \\
    ssl6\_2             & 41          & 152          & 119         & 1          & 0.1          & dll17\_2              & 36          & 82           & 57          & 2            & 0.0          & ssl16                         & 39          & 137          & 119         & -             & 0.0          \\
    ssl6\_3             & 42          & 152          & 107         & 1          & 0.1          & dll18\_3              & 36          & 103          & 78          & 2            & 0.0          & ssl16\_1                      & 39          & 133          & 115         & -             & 0.0          \\
    ssl7                & 41          & 136          & 95          & 1          & 0.1          & dll19                 & 36          & 83           & 63          & 2            & 0.1          & ssl16\_2                      & 47          & 136          & 74          & -             & 0.3          \\
    ssl7\_1             & 40          & 135          & 110         & 1          & 0.1          & dll20\_1              & 36          & 88           & 68          & 2            & 0.1          & ssl16\_3                      & 43          & 133          & 71          & -             & 0.2          \\
    ssl8                & 43          & 166          & 119         & 1          & 0.1          &                       &             &              &             &              &              &                               &             &              &             &               &             
    \\
    \bottomrule
  \end{tabular}
  \end{table*}

\paragraph{Results}

Table~\ref{table:res1} shows {\tool}'s results.
For each model, we list the name, lines of code, the number of nodes that {\tool} determined irrelevant and sliced out, and the number of total AST expression nodes. The last two columns show {\tool}'s resulting ranking of the correct node and its total run time in second.
The 28 italicized \emph{models} contain predicate violations, while the other 124 models contain assertion violations.
{\tool} automatically determines the violation type and switches to the appropriate technique (e.g., using comparator for assertion errors and the unsat analyzer for predicate violations (Section~\ref{sec:approach}).
Finally, the models are listed in sorted order based on their ranking results.


In summary, {\tool} was able to rank the buggy expressions in the top 1 (e.g., the buggy expression is ranked first) for 91 (60\%), top 2 to 5 for 35 (23\%), top 6 to 10 for 10 (34\%), above top 10 for 6 (4\%) out of 152 models.
For 10 models, {\tool} was not able to identify the cause of the errors (e.g., the buggy expression are not in the ranking list), many of which are beyond the reach of {\tool} (e.g., the assertion error is not due to any existing expressions in the model, but rather because the model is ``missing'' some constraints).
Finally, regardless of whether {\tool} succeeds or fails, the tool produces the results almost instantaneously (under a second).

\paragraph{Analysis}
{\tool} was able to locate and rank the buggy expressions in 142/152 models.
Many of these bugs are common errors in which the developer did not consider certain corner cases.
For example, \texttt{stu5} contains the buggy expression \texttt{all n : This.header.*link | n.elem <= n.link.elem} that does not allow any node without link (the fix is changing to 
\texttt{all n : This.header.*link | some n.link => n.elem <= n.link.elem}).
{\tool} successfully recognizes the difference that the last node of the list contains a link to itself in the cex but not in the sat instance, and ranks this expression second; more importantly, it ranks first the subexpression \texttt{n.elem <= n.link.elem}, where the fix is actually needed.
{\tool} also performed especially well on 28 models with violated predicates by analyzing the unsat cores and correctly ranked the buggy expressions first. 

For six models \texttt{bst4\_2}, \texttt{bst16}, \texttt{bst17}, \texttt{bst22\_3}, \texttt{stu9}, and \texttt{stu12\_1}, {\tool} was not able to place the buggy expression within the top 10 (but still within the top 16).
For these models, {\tool} obtains differences that are not directly related to the error, but consistently appear in both the cex and stat instances and therefore confuse {\tool}. 

{\tool} was not able to identify the correct buggy expressions in 10 models, e.g., the resulting ranking list does not contain the buggy expressions.
Most of these bugs are beyond the scope of {\tool} (and fault localization techniques in general).
More specifically, the 9 models \texttt{bst4}, \texttt{bst11}, \texttt{bempl}, \texttt{ctree}, \texttt{dll3\_2}, \texttt{ssl16}, \texttt{ssl16\_1}, \texttt{ssl16\_2}, and \texttt{ssl16\_3} have assertion violations due to \emph{missing} constraints in predicates and thus do not contain buggy expressions to be localized.
For \texttt{other}, {\tool} did not find the "ground truth" buggy expression (the buggy expression does not contain the different relation) but ranked first another expression that could also be modified to fix the error.

\subsection{RQ2: Real-world Case Studies}
\label{sec:casestudies}
The AlloyFL benchmark contains a wide variety of Alloy models and bugs, but they are relatively small ($\approx$50 LoC).
To investigate the scalability of {\tool}, we consider additional case studies on larger and more complex Alloy models.

\begin{table}
\vspace{0.2cm}
  \caption{{\tool}'s results on large complex models}
  \label{tab:realworld}
  \centering
  \begin{tabular}{lr|rrrr}
    \textbf{model}           &  \textbf{loc}       & \textbf{total}     & \textbf{sliced}   &  \textbf{rank}   & \textbf{time(s)}\\
    \midrule
    surgical robots &  200             &     293     & 278  & 2  & 2.3   \\
    \midrule
    android permissions        &  297    & 1138 &  673  & 2    & 5.2 \\
    \midrule
    sll-contains    &  5250             &  5562      & 5510 &  3  &3.0  \\
    count-nodes     &  3791           &  5064      &  2861 & 18  &188.7  \\
    remove-nth      &  5063           &  6336      &  3306 & 12 &1265.0   \\    
    \bottomrule
  \end{tabular}
\end{table}


\paragraph{Surgical Robots} The study in~\cite{robots} uses Alloy to model highly-configurable surgical robots to verify a critical \emph{arm movement safety} property:
the position of the robot arm is in the same position that the surgeon articulates in the control workspace during the surgery procedure and the surgeon is notified if the arm is pushed outside of its physical range.
This property is formulated as an assertion and checked on 15 Alloy models representing 15 types of robot arms using different combinations of hardware and software features.
The study found that 5 models violate the property.

Table~\ref{tab:realworld}, which has the same format as Table~\ref{table:res1}, lists the results.
We use all 5 buggy models (each has about 200 LoC) but list them under one row because they are largely similar and share many common facts and predicates but with different configurable values (one model has a fact that has an AngleLmit set to 3 while another has value 6). The buggy expression is also similar and appears in the same fact.  
For each model, {\tool} ranked the correct buggy expressions in the second place in less than 3 seconds. {\tool} returned two suspicious expressions (1) \texttt{HapticsDisabled in UsedGeomagicTouch.force} and (2) \texttt{some notification : GeomagicTouch.force |  notification = HapticsEnabled}.
Modifying either expression would fix the issue, e.g., changing \texttt{Disable} to \texttt{Enabled} in (1) or \texttt{Enabled} to \texttt{Disabled} in (2).

\paragraph{Android Permissions}
The COVERT project~\cite{covert} uses compositional analysis to model the permissions of Android OS and apps to find inter-app communication vulnerabilities.
The generated Alloy model used in this work does not contain bugs violating assertions, thus we used the MuAlloy mutation tool~\cite{mualloy} to introduce 5 (random) bugs to various predicates in the model: 3 binary operator mutations, 1 unary operator mutation, and 1 variable declaration mutation.

Table~\ref{tab:realworld} shows the result.
{\tool} was able to locate 4 buggy expressions (the unary modification ranked 2nd, the binary operations ranked 3rd, 9th, and 11th), but could not identify the other mutated expression (the variable declaration mutation). However, after manual analysis, we realize that this mutated expression does not contribute to the assertion violation (i.e., {\tool} is \emph{correct} in not identifying it as a fault). 

\paragraph{TACO}
The TACO (Translation of Annotated COde) project~\cite{tacoconf} uses Alloy to verify Java programs with specifications.
TACO automatically converts a Java program annotated with invariants to an Alloy model with an assertion.
If the Java program contains a bug that violates the annotated invariant, then checking the assertion in the Alloy model would provide a counterexample.
We use three different  Alloy models with violated assertions representing three real Java programs from TACO~\cite{tacoconf}:
\texttt{sll-contains} checks if a particular element exists in a linked list;
\texttt{count-nodes} counts the number of a list's nodes; and \texttt{remove-nth} removes the nth element of a list.
These (machine-generated) models are much larger than typical Alloy models ($\approx$5000 to 6000 LoC each).

Table~\ref{tab:realworld} shows the results.
For \texttt{ssl-contains}, {\tool} ranked the buggy expression third within 3 seconds.
This expression helps us locate an error in the original Java program that skips the list header.
The faulty expressions of \texttt{remove-nth} and \texttt{count-nodes} are ranked 12th and 18th, respectively (which are still quite reasonable given the large, $>5000$, number of possible locations).
Note that 
these buggy expressions consist of multiple errors (e.g., having 5 buggy nodes), causing {\tool} to instantiate and analyze combinations of a large number of subexpressions.

Manual analysis on the identified buggy expressions showed that the original Java programs consist of (single) bugs within loops. TACO performs loop unrolling and thus spreads it into multiple bugs in the corresponding Alloy models.




In summary, we found that {\tool} works well on large real-world Alloy models.
While coming up with correct fixes for these models remain nontrivial, {\tool} can help the developers (or automatic program repair tools) quickly locate buggy expressions, which in turn helps understand (and hopefully repair) the actual errors in original models.




\subsection{RQ3: Comparing to AlloyFL}

We compare {\tool} with AlloyFL~\cite{alloyfl}, which to the best of our knowledge, is the only Alloy fault localization technique currently available.
While both tools compute suspicious statements, they are very different in both assumptions and technical approaches.
As discussed in Section~\ref{sec:related}, AlloyFL requires \emph{AUnit tests}~\cite{aunit}, provided by the user or generated from the correct model, and adopts existing fault localization techniques in imperative programs, such as mutation testing and spectrum-based fault localization; in contrast, {\tool} uses violated assertions and relies on counterexamples.

To apply AlloyFL on the 152 benchmark models, we use the best performance configuration and testsuites described in~\cite{alloyfl}. Specifically, we use the AlloyFl$_{hybrid}$ algorithm with Ochiai formula and reuse tests in~\cite{alloyfl} (automatically generated by MuAlloy~\cite{mualloy} as described in~\cite{alloyfl}).

Table~\ref{table:compare} compares the results of {\tool} and AlloyFL.
The two approaches appear to perform similarly, with {\tool} being slightly more accurate. 
Overall, {\tool} outperforms AlloyFL, where on average the  buggy expressions ranked 2nd and 3rd by {\tool} and AlloyFL, respectively. Also, in top 1 ranking {\tool} performs much better compared to AlloyFL (91 over 76 models). 
Moreover, {\tool} is much faster, where the average analysis time is far less than a second for {\tool}, it takes over 30 seconds for AlloyFL to analyze the same specifications.

We were not able to apply AlloyFL to the models in Section~\ref{sec:casestudies} because MuAlloy~\cite{mualloy}, which is used to generate AUnit tests for AlloyFL, does not work with these models (e.g., mostly caused by unhandled Alloy operators). 
This is not a weakness of AlloyFL, but it suggests that it is not trivial to generate tests from existing Alloy models automatically.




\begin{table}
  \centering
  \vspace{0.2cm}
  \caption{Comparision with AlloyFL.}
  \label{table:compare}
  \begin{tabular}{c | c c c c c | c c}
    \toprule
    &\multicolumn{4}{c}{\textbf{top}}&&\multicolumn{2}{c}{\textbf{avg}}\\
    \textbf{tool} & 1& 5 & 10 & $>$ 10 & \textbf{failed} & \textbf{rank}& \textbf{time(s)} \\
    \midrule
    {\tool}&91&126&136&6&10&2.4&0.2\\
    AlloyFL&76&128&137&8&7&3.1&32.4\\
    \bottomrule
  \end{tabular}
\end{table}

\subsection{Threats to Validity}
We assume no fault in data type (sig's) and field declarations, which may limit the usage of {\tool}.
However, none of the benchmark models we used has bugs at these locations.
Moreover, we could always translate constraints for sig and field to facts. For example, \texttt{one sig A} could be translated to \texttt{sig A; fact\{one A\}}.

The models in the AlloyFL benchmark are collected from graduate student's homework and relatively small.
Thus, they may not represent faulty Alloy models in the real world. We also evaluate {\tool} with large Alloy models, written by experienced Alloy developer (the surgical robot models and Android permissions model) or generated by an automatic tool (TACO) and 
show that {\tool} performs well on these models (Section~\ref{sec:casestudies}).

We manually create assertions for models that do not have assertions. Thus, our assertions might be inaccurate and not as intended.
However, for other models with assertions (e.g., those in the AlloyFL benchmark and all the case studies), we use those assertions directly and {\tool} output similar results.




\section{Related Work}
\label{sec:related}

{\tool} is related to AlloyFL~\cite{alloyfl}, which 
adopts spectrum-based fault localization~\cite{spfl1, spfl2, spfl3} and mutation-based techniques~\cite{mut1, mut2} from imperative languages.
Given AUnit tests~\cite{aunit} labeled with should-pass or should-fail, AlloyFL computes a suspicious score for expression by mutating and giving it a higher score if the mutation increases the number of should-pass tests pass and the number of should-fail tests fail.
AlloyFL uses MuAlloy~\cite{mualloy} to automatically generate tests.
However, MuAlloy requires the correct Alloy model to generate these tests.
{\tool} does not require tests and instead uses assertions, which are commonly used in Alloy.

The generation of similar instances can be viewed as a model exploration problem~\cite{modelfind}. Bordeaux~\cite{bordeaux} uses Alloy\* to find pairs of SAT/UNSAT instances with minimum relative distances. In contrast, {\tool} reduces the generation of an instance as close as possible to the identified counterexamples into a partial max sat problem and solves it using a PMAX-SAT solver. 

Amalgam~\cite{whyandwhynot} explains why some tuples of a relation do or do not appear in certain instances. A user would manually select a tuple to add or delete, and Amalgam tries to explain why they can or cannot do so (typically the reason for counterexamples is due to the assertions). {\tool} instead automatically identifies why a counterexample fails and finds locations that relate to this violation.


Many fault localization techniques have been developed for imperative languages.
Spectrum-based techniques~\cite{spfl1,spfl2,spfl3,spfl4,spfl5,tarantula,spfl6,spfl7} identify faulty statements by comparing passing and failing test executions.
SAT-based techniques~\cite{sat1, sat2} convert the fault location problem into an SAT problem.
Statistic-based methods~\cite{stat1, stat2} collect statistical information from test executions to locate errors. Feedback-based techniques~\cite{feed1, feed2} interactively locates error by getting feedback from the user.
Delta debugging~\cite{delta1, delta2} identifies code changes responsible for test failure.
There are also works on minimizing differences in inputs based on the assumption that similar inputs would lead to similar runs~\cite{diff1, diff2, diff3}.
Program slicing~\cite{slice1, slice2, slice3} has also been used to aid debugging.


Model-based diagnosis (MBD) approaches identify faulty components of a system based on abnormal behaviors.
Griesmayer~\cite{mdb3} applied MBD to localizing fault in imperative programs using model checker. Marques-Silva~\cite{mdb2} converted the MBD problem into a MAXSAT problem to find the minimal diagnosis, where the system description is encoded as the hard clauses and the not abnormal predicates as the soft clauses.
There has also been another similar line work in pinpointing axioms in description logic~\cite{axoimpinpoint, dl}.


\section{Conclusion and Future Work}
We introduce a new fault localization approach for declarative models written Alloy.
Our insight is that Alloy expressions that likely cause an assertion violation can be obtained by analyzing the counterexamples, unsat cores, and satisfying instances from the Alloy Analyzer.
We present {\tool}, a tool that implements these ideas to compute and rank suspicious expressions causing an assertion violation in an Alloy model.
{\tool} uses a PMAX-SAT solver to find satisfying instances similar to counterexamples generated by the Alloy Analyzer, analyzes satisfying instances and counterexamples to locate suspicious expressions, analyzes subexpressions to achieve a finer-grain level of localization granularity, and uses unsat cores to help identify conflicting expressions. 
Preliminary results on existing Alloy benchmarks and large, real-world benchmarks show that {\tool} is effective in finding accurate expressions causing errors. 
We believe that {\tool} takes an important step in finding bugs in Alloy and exposes opportunities for researchers to exploit new debugging techniques for Alloy.

Currently, we are improving the accuracy and efficiency of {\tool}. Specifically, instead of using a default number of instance pairs, we can search for instances incrementally until the algorithm converges.
We are also exploring new approaches to effectively integrate {\tool} with automatic Alloy repair techniques. Preliminary results from the recent BeAFix work~\cite{brida2021bounded} shows that {\tool} accurately identifies faults in Alloy specifications, which in turn helps BeAFix automatically analyze and repair those specifications.



\section{Data Availability}

We make {\tool} and all research artifacts, models, and experimental data reported in the paper available to the research and education community~\cite{BFA}.

\section*{Acknowledgment}
We thank the anonymous reviewers for helpful comments.
This work was supported in part by awards W911NF-19-1-0054 from the Army Research Office; CCF-1948536, CCF-1755890, and CCF-1618132 from the National Science Foundation; and  PICT 2016-1384, 2017-1979 and 2017-2622 from the Argentine National Agency of Scientific and Technological Promotion (ANPCyT).
\balance
\bibliographystyle{IEEEtran}
\bibliography{paper}

\end{document}